\documentclass[amssymb,amsmath,aps,showpacs,floatfix,nofootinbib,showpacs,12pt]{revtex4}
\usepackage{amssymb}
\usepackage{graphicx}
\usepackage{color}
\usepackage{soul}
\usepackage{latexsym}

\newcommand{\lsim}{\lesssim}
\newcommand{\gsim}{\gtrsim}

\def\lsim{\mathrel{\raise.3ex\hbox{$<$\kern-.75em\lower1ex\hbox{$\sim$}}}}
\def\gsim{\mathrel{\raise.3ex\hbox{$>$\kern-.75em\lower1ex\hbox{$\sim$}}}}

\def\beq{\begin{equation}}
\def\eeq{\end{equation}}
\def\beqn{\begin{eqnarray}}
\def\eeqn{\end{eqnarray}}
\def\bea{\begin{eqnarray}}
\def\eea{\end{eqnarray}}
\def\be{\begin{equation}}
\def\ee{\end{equation}}
\newcommand{\fslash}[1]{{#1 \kern -0.7em/ \kern 0.1em}}

\begin{document}

\voffset 1.25cm

\title{
Neutrino Signals from Solar Neutralino Annihilations in Anomaly
Mediated Supersymmetry Breaking Model}

\author{Jia Liu, Peng-fei Yin and Shou-hua Zhu}
\affiliation{Institute of Theoretical Physics \& State Key
Laboratory of Nuclear Physics and Technology, School of Physics,
Peking University, Beijing 100871, China}

\date{\today}

\begin{abstract}

The lightest neutralino, as the dark matter candidate, can be
gravitationally  captured by the Sun. In this paper, we studied the
high energy neutrino signals from solar neutralino annihilations in
the core of the Sun in the anomaly mediated supersymmetry (SUSY)
breaking (AMSB) model. Based on the event-by-event monte carlo
simulation code WimpSim, we studied the detailed energy and angular
spectrum of the final muons at large neutrino telescope IceCube.
More precisely we simulated the processes since the production of
neutrino via neutralino annihilation in the core of the Sun,
neutrino propagation from the Sun to the Earth, as well as the
converting processes from neutrino to muon. Our results showed that
in the AMSB model it is possible to observe the energetic muons at
IceCube, provided that the lightest neutralio has relatively large
higgsino component, as a rule of thumb $ N_{13}^2 + N_{14}^2  > 4\%$
or equivalently $ \sigma _{SD}
> 10^{ - 5} pb$. Especially, for our favorable parameters the signal
annual events can reach 102 and the statistical significance can
reach more than 20. We pointed out that the energy spectrum of muons
may be used to distinguish among the AMSB model and other SUSY
breaking scenarios.

\end{abstract}

\pacs{95.35.+d, 12.60.Jv, 13.15.+g, 95.55.Vj}


\maketitle

\section{Introduction}

Over the past several years, the picture that our universe is not
baryon dominant is well established. Based on the observations of
Wilkinson Microwave Anisotropy Probe (WAMP) and other experiments,
the relic dark matter density is fixed to be $0.085< \Omega_{DM}h^2<
0.119$ within $2\sigma$ uncertainty \cite{WMAP}. In order to account
for the dark matter, physics beyond the standard model (SM) is
usually required. Among these new physics the weakly interacting
massive particle (WIMP) is one of the popular candidate for the dark
matter. One possible WIMP is the lightest supersymmetric (SUSY)
particle (LSP) in SUSY theory with R-parity conservation
\cite{JKG95BHS04}. Usually there are three possible LSPs, namely
sneutrino, gravitino and neutralino. The lightest sneutrino has been
largely ruled out by direct searches \cite{sneutrino}, and the
gravitino is difficult to be detected due to its very weak
interactions with ordinary matter \cite{FRT03}. Thus the lightest
neutralino, which is combination of gaugino and higgsino, is the
most well-motivated candidate for dark matter \cite{neutralino}.

SUSY breaking mechanism plays an important role to investigate the
nature of dark matter. Firstly, in order to be consistent with
experimental observations, SUSY must be broken at weak scale. As a
consequence, the mass of dark matter is partially determined by the
breaking terms.  Secondly, we need higher scale SUSY breaking
mechanism to reduce the number of free parameters at low energy. As
an example there are more than 100 parameters in minimal
supersymmetric standard model(MSSM). Thirdly, the properties of the
LSP depend on the details of SUSY breaking mechanism. For example in
minimal supergravity (mSUGRA) model, the main component of the
lightest neutralino can be the bino in 'bulk region', while in
'focus point region', the lightest neutralino can have larger
higgsino content \cite{CCN98, FMM, FW05, BKPU05, FMW}. Here 'bulk
region' and 'focus point region' are determined by different SUSY
breaking parameters. Thus it is interesting to explore (i) the
predictions for different SUSY breaking mechanisms; and (ii) whether
the detection of dark matter can distinguish among the different
SUSY breaking mechanisms.

While in literature the mSUGRA model was widely investigated, in
this paper we are interested in the properties of neutralino (as the
dark matter particle) and its detection in the scenario of the
anomaly mediated SUSY breaking (AMSB)\cite{RS98,GLMR98}. Many
predictions in AMSB models are different from those in usual mSUGRA
models and other SUSY breaking scenarios. One feature of AMSB model
is the special gaugino mass relation calculated by $\beta$ function,
$M_{1}:M_{2}:|M_{3}|=2.8:1:7.1$ \cite{GGW99}, which implies that the
lightest chargino pair and neutralino comprise a nearly
mass-degenerate triplet of winos over most of the MSSM parameter
space. Moreover the lightest neutralino is mostly wino-like. Such
kind of neutralino usually has larger cross section compared to that
in mSUGRA model. As a consequence, if it is the dark matter, it
should be produced in non-thermal processes \cite{GGW99,MR99}, or
should be very heavy, generally larger than $2\ TeV$
\cite{W04,CDKR06,HMNSS06}.

The methods to detect dark matter can be classified into direct and
indirect ones. The former methods detect dark matter by measuring
the recoil of nucleus. The latter ones detect the final stable
particles, including neutrinos, antiprotons, positrons, antinuclei
and photons, which are produced by dark matter annihilation. In this
paper we will focus on the observation by neutrino telescopes which
detect the high energy neutrinos from dark matter annihilations (for
some recent works, see
\cite{FMW,highE,BKS01BDK02,BHHK02,HK02,HS05,iceneutrino,LE04,CFMSSV05,BKST07,LW07}).
The non-relativistic dark matter would be trapped by Sun or Earth
via gravitational force. The captured dark matter can annihilate
into SM particles, including neutrinos, in the center of Sun or
Earth\cite{PS85, neutrino,earlywork}. The resulting neutrinos
interact with usual matter feebly, thus it is possible for us to
detect them after a long propagation. Many neutrino telescopes for
high energy neutrino have been built or are under construction such
as IceCube \cite{icecube,Aetal04}, Super-Kamiokande \cite{superk},
AMANDA \cite{amanda} and ANTARES \cite{antares} etc. In this paper
we will focus on the neutrino detection at IceCube. The IceCube
detector now is being built in the south pole and will be completed
in 2010 \cite{icecube,Aetal04}. The scale of the IceCube is in
square kilometers and it can detect neutrinos with threshold energy
of $50\ GeV$\cite{icecube}.

In this paper, we will discuss neutrino signals from neutralino
annihilations in the Sun in AMSB scenario. The final muon events
rate at the IceCube detector can be determined by \cite{E93}:
\begin{equation}
\Gamma_{events}=\Gamma_{ann}A_{eff}\sum_{i}Br_{i}\Phi_{\mu
i}\label{eq0}
\end{equation}
where $\Gamma_{ann}$ is neutralino's annihilation rate in the Sun,
$A_{eff}$ is detector's actual effective detected area, $Br_{i}$ is
the branching fraction of i-th annihilation channel, and
$\Phi_{\mu}$ is the $\mu$ flux over detector threshold per unit area
and parent pair. Our study shows that for the Sun there exit the
parameter space which can induce the large neutralino capture rate
and thus the large neutralino annihilation rate $\Gamma_{ann}$. In
this case, most neutrinos come from $W$ decay and the $W$s are pair
produced via neutralino annihilations. Such neutralino is wino-like.
Recently the authors of Ref. \cite{BKST07} pointed out that if the
spin effects of neutralino annihilation are included, the energy
spectrum of neutrinos in the Sun will be different. Thus in our
study we include such kind of spin effects. We utilize Calchep
\cite{comphep,calchep} to generate events of neutralino annihilation
based on exact matrix elements calculation. We then import these
events into Monte Carlo simulation program - WimpSim \cite{wimpsim}
which performs a thorough analysis for neutrinos propagation from
the Sun to detector at the surface of the Earth \cite{BEO07}.  The
program contains some important features similar to those of Pythia
\cite{pythia}, Darksusy \cite{darksusy} and nusigma \cite{nusigma}.
Thus WimpSim can analyze details of neutrino propagation and
detection including both neutrino oscillations and interactions
\cite{BEG, E93,E97,BEO07}. The detail Monte Carlo simulations allow
us tracing every neutrino event and determining final $\mu$ signals
with the concrete energy and angle information. At the same time, we
will consider the real environment of IceCube detector
\cite{Aetal04,GHM05}, and obtain the realistic final muon spectrum.

This paper is organized as following. In  section II we will briefly
discuss some features of AMSB model and the properties of the
neutralino. In section III, we perform the parameter space scan to
determine different neutralino capture rates for the Sun. In section
IV, we explore all important neutralino annihilation channels, then
calculate the spectrum of neutrinos from the production of
neutralino annihilation. In section V, we investigate the
propagation of neutrinos from center of the Sun to the Earth. In
section VI, the final spectrum of $\mu$ is presented. The last
section contains our conclusions and discussions.

\section{Neutralino in the AMSB model}

Anomaly-mediated contributions to SUSY breaking usually appear in
supergravity theory, but they are loop suppressed compared to those
of gravity-mediated. However the latter contributions are assumed to
be zero in the AMSB scenario. In this scenario there are no direct
interaction between hidden and visible sectors \cite{RS98,GLMR98}.
Thus the masses of neutralino and slepton are zero at tree level.
The soft SUSY breaking terms are related to the super-conformal
anomaly, and they appear at loop level. In the hidden sector, an
auxiliary field as a super-gravity ground can be thought as the only
origin to break SUSY. In order to have a conformal lagrangian, a
compensator superfield $\phi$ is introduced \cite{RS98,GLMR98,PR99}.
We can assume that $\phi$ get non zero value as
\begin{equation}
\phi=1 + <F_{\phi}>\theta^{2},
\end{equation}
then expand $\phi$ in background value $<F_{\phi}>=-m_{3/2}$
($m_{3/2}$ is related to the mass of gravitino) and perform
calculation with re-scaling the coupling of matter field. The soft
SUSY breaking terms related to gauginos and sleptons are calculated
to be \cite{GGW99}:
\begin{eqnarray}
M_{\lambda}&=&\frac{\beta_{g}}{g}m_{3/2} \label{eq2}\\
m_{\bar{Q}}^2&=&-\frac{1}{4}(\frac{\partial\gamma}{\partial g}\beta_g+\frac{\partial\gamma}{\partial y}\gamma_y)m_{3/2}^2 \label{eq1}\\
A_y&=&-\frac{\beta_y}{y}m_{3/2},
\end{eqnarray}
where $\gamma$, $\beta$ are 
functions defined as
\begin{equation}
\gamma\equiv\frac{d\ln Z}{d\ln\mu},\,\,\, \beta_g\equiv\frac{dg}{d
\ln\mu},\,\,\, \beta_y\equiv\frac{dy}{d \ln\mu}.
\end{equation}
From these formulas we can see that the masses of SUSY particles in
AMSB scenario are determined by $m_{3/2}$ and evolutions of gauge
and Yukawa couplings. It is interesting to note that the mass of
slepton in the Eqn.(\ref{eq1}) is negative.  In literature several
solutions are proposed \cite{PR99,KK00,CL01,MSS08}.
Phenomenologically one can simply add a universal positive mass
terms $m_0^2$ to the right hand of Eqn.(\ref{eq1})
\cite{GGW99,FM99}. The resulting model is dubbed as minimal AMSB
(mAMSB) model which only depends on
\begin{equation}
m_{3/2}, \, m_0, \, \tan\beta, \,sign(\mu)
\end{equation}
as the extra free parameters.

The mAMSB model is highly predictive. One of the important
predictions is the relationship between gaugino masses as
$M_{1}:M_{2}:|M_{3}|=2.8:1:7.1$ at the low energy scale, which
implies that the lightest neutralino is mostly wino-like over most
of the MSSM parameter space. It is known that in MSSM the neutralino
is a linear combination of gauginos and higgsinos, and the lightest
neutralino is given as
\begin{equation}
{\widetilde\chi _1^0 }
=N_{11}\tilde{B}+N_{12}\tilde{W}+N_{13}\tilde{H_d}+N_{14}\tilde{H_u}.
\end{equation}
On the contrary the lightest one is bino-like over most of the
mSUGRA parameter space with GUT assumption.

\begin{figure}[h]
\vspace*{-.03in} \centering
\includegraphics[width=4.0in,angle=0]{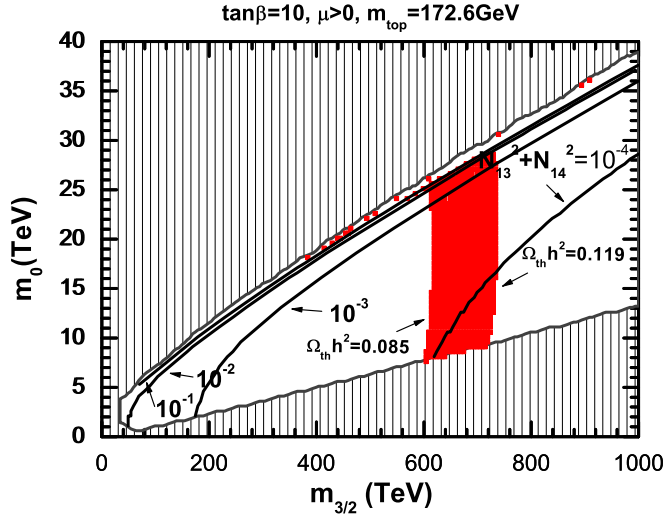}
\vspace*{-.03in} \caption{ The allowed points which can induce
correct thermal relic density $0.085 < \Omega _{th}h^2 < 0.119$ in
the plane of $m_{3/2}$ and $m_0$. Curves of $N_{13}^2 + N_{14}^2$
(as a measure for the lightest neutralino Higgsino components) are
also shown. The left-top region is excluded because it can not
radiatively lead to EWSB, and the right-bottom region is excluded
because it has tachyon. The allowed points are generated by
MicrOMEGAs \cite{micromegas}. \vspace*{-.1in}} \label{relic}
\end{figure}

In Fig.~\ref{relic} we show the points  which can result in the
correct thermal relic density in $m_{3/2}$ and $m_0$ plane. We
utilize Suspect code \cite{suspect} to scan parameter space in mAMSB
model. For the case of $N_{13}^2+N_{14}^2 \lesssim 0.1$, the relic
density can be approximately given as \cite{GR04}
\begin{equation}
\Omega_{th}h^2 \simeq 0.02(\frac{M_2}{1TeV})^2.
\end{equation}
Such kind of lightest neutralinos usually has large annihilation
cross sections in the thermodynamic equilibrium of early universe,
therefore it is not easy to induce the correct relic density.
However one can still obtain the correct density if lifting the mass
of neutralino to be as heavy as $1.9\sim 2.3$ TeV \cite{CDKR06,W04}.
If wino-like neutralino with $SU(2)_L$ charge is much heavier than
the weak gauge boson, the weak interaction is a long-distance force
for non-relativistic two-bodies states of such particles. If this
non-perturbative effect (namely Sommerfeld enhancement) of the dark
matter at the freeze-out temperature is taken into account, the
abundance can be reduced by $\sim 50\%$. Thus the neutralino mass
would be as heavy as $2.7\sim 3.0$ TeV\cite{HMNSS06}. Such heavy
neutralino may be detected through dark matter search experiments
\cite{CDKR06,HMNS0412,HMSS05}, but it is difficult to detect and
study it at the LHC and the planning ILC. For the case of
$N_{13}^2+N_{14}^2 > 0.1$, the light neutralino can also induce the
correct density for a tiny region of parameter space, which lies in
the vicinity of boundary of the electroweak symmetry breaking
(EWSB). Here the $|\mu|$ decreases and Higgsino components become
significant.

Now that the light neutralino hardly accounts for thermal relic
density in the mAMSB model, another approach is proposed, i.e. the
neutralino is not mainly from thermal production. Some authors
discussed the mechanism that the LSP is produced by the decay of
gravitinos \cite{GGW99,MR99} (or moduli fields \cite{MR99}) in the
AMSB model. From Eqn.(\ref{eq2}) we can see that neutralino is much
lighter than that of gravitino. If the gravitinos have a relatively
short lifetime, this mechanism will not destroy the success of
big-bang nucleosynthesis (BBN)  \cite{MR99}. In this scenario, the
light LSP with large wino component can be the good candidate for
dark matter. However it should be pointed out that although this
kind of LSP with large cross section could be produced in
non-thermal process, it is also constrained by cosmological
observations. For example, in the center of our galaxy, we may
observe the signals of the dark matter annihilation
\cite{BE98HS04PU04,HD02,YHBA07} due to their large cross sections.
Recently, Ref. \cite{H08} reported that if assuming an excess of
microwave emission in the inner Milky Way is due to dark matter
annihilation, the synchrotron measurements give constraint to the
annihilation rate of neutralino. The constraint depends on the dark
matter distribution profile. More stringent constraint has been
reported by Ref. \cite{J04}. The residual annihilation of dark
matter would produce $^6$Li during the BBN. Then the observations of
abundance of $^6$Li could constrain the dark matter annihilation
rate. The low mass wino-like dark matter below $250$ GeV in AMSB
model has been ruled out \cite{J04}. It is still possible to detect
heavier wino-like neutralino at the high energy collider
\cite{detamsbcol} and non-accelerator dark matter search experiments
\cite{MR99,U01,HW04,HMNS0407}.

\begin{figure}[h]
\vspace*{-.03in} \centering
\includegraphics[width=4.0in,angle=0]{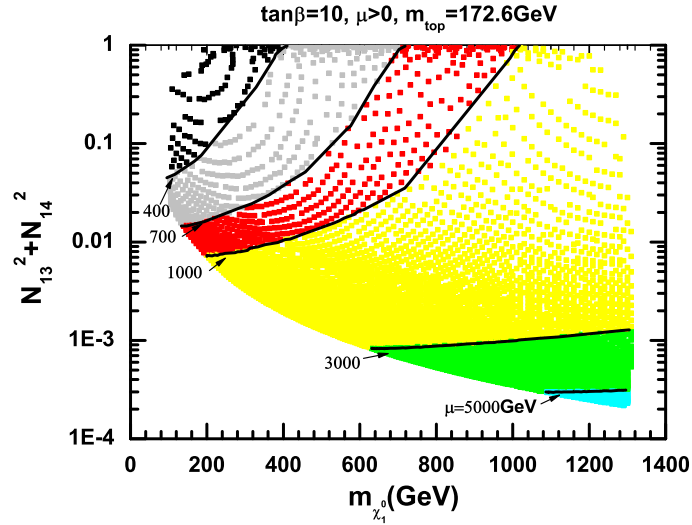}
\vspace*{-.03in} \caption{  Points  which can induce relic density
lower than 0.119 in the plane of lightest neutralino mass and
$N_{13}^2+N_{14}^2$. Here $ 0 < m_0 < 20TeV $ and $0 < m_{3/2}  <
400TeV$.  The curves with different $\mu$ at 5000, 3000, 1000, 700,
400 GeV are also plotted. We can see that the light neutralino with
small $\mu$ is easy to get large Higgsino component.
 \vspace*{-.1in}}
\label{hig}
\end{figure}

The general relation between $\mu$ and other parameters in SUSY at
tree level is given by
\begin{equation}
\mu^2=\frac{m_{H_d}^2-m_{H_u}^2\tan^2\beta}{\tan^2\beta-1}-\frac{1}{2}m_Z^2.
\end{equation}
As discussion in Ref. \cite{FM99}, pure anomaly-mediated value of
$m_{H_u}^2$ is renormaliztion group (RG) invariant, and non
anomaly-mediated contribution of $\delta m_{H_u}^2$ would be even
zero for some parameters. This focusing behavior near the weak scale
makes $m_{H_u}^2$ only lightly dependent on ultra-violet (UV)
boundary and would give suitable small $\mu$ for naturalness
requirement. At the same time it will lead to considerable higgsino
component of the LSP, which is similar as that
 in 'focus point region' in mSUGRA model \cite{CCN98,
FMM, FW05, BKPU05, FMW}. In Fig. \ref{relic} and Fig. \ref{hig}, the
parameter scan was done by using MicrOMEGAs \cite{micromegas} and
Suspect\cite{suspect} with parameters $ \tan \beta  = 10, \mu  >
0,m_{top}  = 172.6GeV$ \cite{CDFD0}. In Fig.\ref{relic}, the
left-top region is excluded because the radiative EWSB condition
fails. With the decrement of $|\mu|$, the Higgsino composition
becomes larger near the boundary. In this parameter region, the
lightest neutralino could be partially non-thermal-produced. At the
same time, some points can even directly satisfy thermal relic
density due to relative small wino content. In Fig. \ref{hig} we
show the points which can induce relic density lower than 0.119 in
the plane of $m_{\tilde{\chi}_1^0}$ and $N_{13}^2+N_{14}^2$. Here we
choose $ 0 < m_0 < 20$ TeV  and $0 < m_{3/2}  < 400$ TeV, which is
smaller than those in Fig. \ref{relic}. The reason is that the heavy
neutralino, which corresponds to large $m_0$ and $m_{3/2}$, usually
have small spin-dependent cross-section $ \sigma_{SD}$ (see Fig.
\ref{SDvM} below), consequently low capture rate in the Sun. We
impose the constraints as
 $ m_h > 114.4GeV$ \cite{ADLO03}, relic density
$ \Omega h^2 < 0.119$ \cite{WMAP}, $ \Delta \rho  < 2 \times 10^{ -
3}$, $  B(b \to s\gamma ) = (355 \pm 26) \times 10^{ - 6}
$\cite{HFAG}  and $  \Delta a_\mu (Exp - SM) = (26.1 \pm 9.4) \times
10^{ - 10} $ \cite{g-2} within $ 3\sigma$ uncertainty. Note that in
scattered plots in Fig. \ref{sdvall}, Fig. \ref{SDvM} and Fig.
\ref{br}, we also choose $ 0 < m_0 < 20TeV $, $0 < m_{3/2} < 400TeV$
and with the same parameters and constraints as in Fig. \ref{hig}.
In our chosen parameter space, the relic density is well under upper
limit of WMAP. A few words on the constraint from $ a_\mu$ is worth
to mention. Recently, the authors of Ref. \cite{HMNT07} updated SM
prediction of $ a_\mu$ which has a smaller uncertainty and
corresponds to a $ 3.4\sigma$ deviation from the measured value in
Ref. \cite{g-2}. In mAMSB model, parameter space with large $ m_0$
and $ m_{3/2}$ is heavily constrained because the SUSY contributions
are suppressed by heavy superparticle masses. In mSUGRA model, for
the same reason the parameter space with large $ m_0$ and $ m_{1/2}$
is also severely constrained.

\section{Capture and annihilation rate of dark matter in the AMSB model}

There are dark matter in our galactic halo, and the dark matter
particles will scatter off nucleus in astrophysical objects such as
the Sun and the Earth and be gravitationally trapped in them
\cite{PS85}. Once captured, dark matter particles will have a larger
annihilation rate due to the larger density and will produce the
energetic SM particles. These SM particles are mostly absorbed by
the matter except neutrinos which interact feebly with ordinary
matter. The evolution of dark matter number in the objects can be
written as
\begin{equation}
\dot{N}=C_ \odot -C_A N^2.
\end{equation}
Here $C_ \odot$ is the capture rate and $C_A \equiv < \sigma v
> /V_{eff}$ is thermally averaged annihilation cross section per
volume. The annihilation rate can be solved as
\begin{equation}
\Gamma=\frac{1}{2}C_AN^2=\frac{1}{2}C_\odot \tanh^2(\sqrt{C_ \odot
C_A}t),
\end{equation}
where $t$ is the age of this system and $\frac{1}{2}$ is due to the
two LSPs annihilation. For the Sun, $\sqrt{C_ \odot C_A}t\gg1$ leads
an approximation of above result as $\Gamma=\frac{1}{2}C_\odot $.
 The capture rate for the Sun is then given
as \cite{G92}
\begin{equation}
C_ \odot   = 3.4 \times 10^{20} s^{ - 1} \frac{{\rho _{local}
}}{{0.3GeV/cm^3 }}(\frac{{270km/s}}{{v_{local} }})(\frac{{\sigma
_{SD}^H  + \sigma _{SI}^H  + 0.07\sigma _{SI}^{He} }}{{10^{ - 6}
pb}})(\frac{{100GeV}}{{ m_{\widetilde\chi _1^0 }
  }})^2,
\end{equation}
where $\rho_{local}$ is the local mass density of dark matter,  and
$v_{local}$ is velocity of these particles. In our numerical
evaluation we adopt the values $\rho_{local}=0.3\ GeV/cm^3$ and $
v_{local} =220\ km/s$ \cite{PDG}. Here $\sigma_{SD}$ and
$\sigma_{SI}$ are spin-dependent and spin-independent scattering
cross sections of LSP with nucleus respectively.

In the capture process, LSPs lose their energy by scattering off the
nucleus in the Sun and then trapped by gravity. The capture rate
depends on the scattering cross section of LSP with nucleus. The
spin-independent scattering cross section  $ \sigma _{SI} $ has been
strictly constrained by direct detection on the Earth. For example
XENON10 set an upper limit for the WIMP-nucleon spin-independent
cross section of $ 8.8 \times 10^{ - 8} pb$ for a WIMP mass of $100
GeV$, and $ 4.5 \times 10^{ - 8} pb $ for a WIMP mass of $30 GeV$ at
90\% confidence level \cite{AA08}.

The cross section for scattering of a neutralino off of a proton via
spin-dependent interaction is \cite{K91,RABGMR93},
\begin{equation}
\sigma_{SD}= \frac{{32m_{\widetilde\chi _0^1 }^2 m_p^2 G_F^2 }}{{\pi
(m_{\widetilde\chi _0^1 }  + m_p )^2 }}J(J + 1)[\sum\limits_{u,d,s}
{A'_q \Delta q} ]^2,
\end{equation}
where
\begin{eqnarray}
 A'_q &= & \frac{1}{2}T_{3L}^q \left( |N_{13}|^2  -|N_{14}|^2 \right) \nonumber \\
 && - \frac{{m_W^2 }}{{m_{\widetilde\chi _0^1 }^2  - m_{\widetilde q}^2 }}
  \left\{ \frac{{m_q^2 d_q^2 }}{{2m_W^2 }}+ \left[T_{3L}^q N_{12}-\tan \theta _W (T_{3L}^q  - e_q )N_{11}\right]^2
   +\tan ^2 \theta _W e_q^2 N_{11}^2  \right\}.
     \label{sigsd}
\end{eqnarray}
Here $ m_q$ is the quark mass, $ d_q  =  - N_{13} /\cos \beta$ for
down-type quarks, $ d_q  =  - N_{14} /\sin \beta$ for up-type
quarks,  $ T_{3L}^q$ is the weak isospin of the quark  and $ e_q$ is
quark electric charge. The quantity $ \Delta q$ measures the
fraction of the nucleon spin carried by the quark. The values are
taken as $ \Delta u = 0.78$, $ \Delta d =  - 0.5$ and $ \Delta s = -
0.16$ \cite{K91,RABGMR93,EK93,BB98}. The first part of the Eqn.
\ref{sigsd} corresponds to the process of exchanging Z boson,
 and the second part of exchanging
squark. For heavy squark, the second term can be safely neglected.

\begin{figure}[h]
\vspace*{-.03in} \centering
\includegraphics[width=4.0in,angle=0]{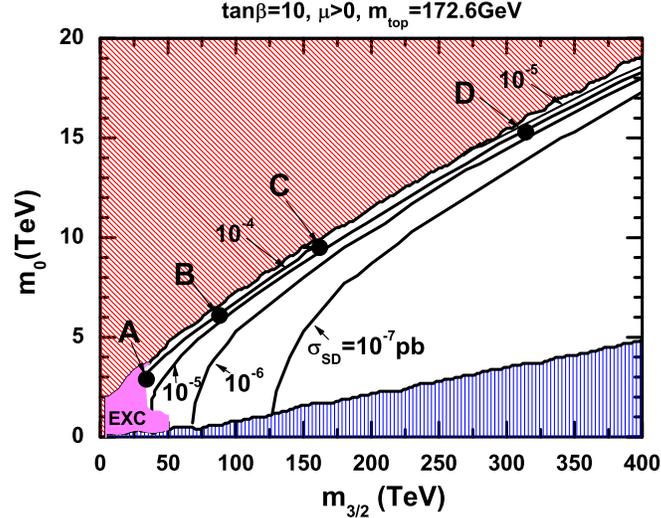}
\vspace*{-.03in} \caption{Contours of $\sigma_{SD}$ in the
($m_{3/2}, m_0$) plane. The constraints are the same as those in
Fig.~\ref{hig}. Four benchmark points A, B, C and D, which are used
in Tab.~\ref{table1} and
 Tab.~\ref{table2}, are
also depicted.
 \vspace*{-.1in}}
\label{sdvall}
\end{figure}

\begin{figure}[h]
\vspace*{-.03in} \centering
\includegraphics[width=4.0in,angle=0]{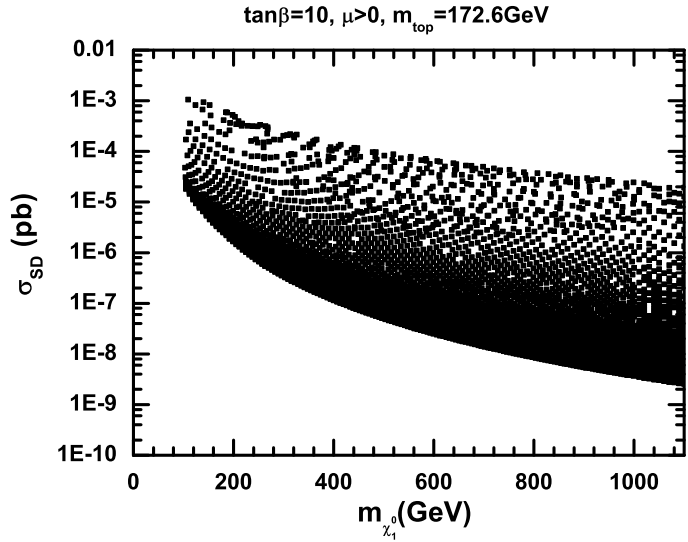}
\vspace*{-.03in} \caption{Allowed parameter points in the plane of
lightest neutralino mass and $ \sigma _{SD}$. The parameters
$\tan\beta, sign(\mu), m_{top}$ are the same as those in
Fig.~\ref{sdvall}. The constraints are the same as those in
Fig.~\ref{hig}.
 \vspace*{-.1in}}
\label{SDvM}
\end{figure}

The cross sections of $ \sigma _{SD}$ in AMSB model are plotted in
Fig.~\ref{sdvall} and Fig.~\ref{SDvM}. In these two figures the
largest neutralino mass is $ 1.1TeV$. For heavier neutralino the
capture rate is rather low because $\sigma_{SD}$ is smaller and it
is also suppressed by its mass.
 In Fig. \ref{SDvM}, the $\sigma _{SD}$ varies from $10^{ - 9}pb$ to
 $10^{ - 3}pb$. Note that in Fig. \ref{sdvall} the lightest
 neutralino,
 which lies in the vicinity of
 EWSB boundary, has higher
 higgsino fraction thus larger cross section $\sigma _{SD}$.
 In Fig. \ref{sdvall}, the lower left portion is excluded by direct
 accelerator limits on sparticle masses, light Higgs mass
 constraint.

\section{Production and propagation of neutrinos}

\subsection{Production of neutrinos in the Sun}

\begin{figure}[h]
\vspace*{-.03in} \centering
\includegraphics[width=3.3in,angle=0]{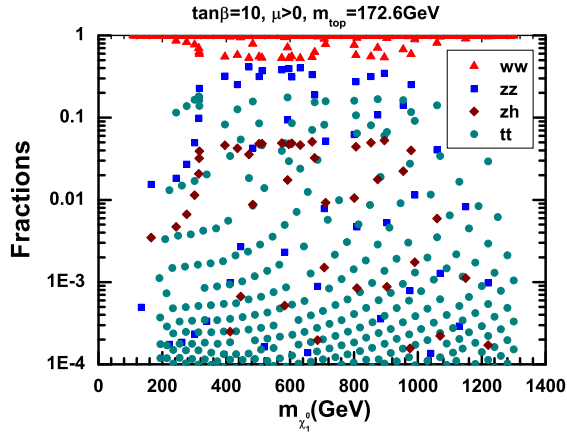}
\vspace*{-.03in} \caption{Fractions for neutralinos annihilate into
$WW$, $ZZ$, $ZH$ and $t\bar t$ as a function of neutralino mass.
 \vspace*{-.1in}}
\label{br}
\end{figure}

The dark matter annihilation in the Sun is calculated in the static
limit, i.e. the dark matter is assumed to be at rest, because the
velocity of these particles is about $220km/s$ \cite{PDG} in the
solar system. The neutrinos can be directly produced from the
lightest neutralinos $ \widetilde\chi _0^1$ annihilation. However
the neutrino flux is usually negligible compared to those from $
\widetilde\chi _0^1$ annihilation to $ W^+ W^-, ZZ, Zh, f\overline
f$. In this paper the direct neutrino production from $
\widetilde\chi _0^1$ annihilation is not included. In mSUGRA model,
usually contributions from each channel to neutrino flux play the
same important role. However in AMSB model, $ \widetilde\chi _0^1$s
mainly annihilate to $ W^ + W^ -$ which is usually 10-1000 times
greater than other modes. The reason is that $ \widetilde\chi _0^1$
is mostly constituted by wino and chargino is also composed nearly
of charged wino, thus the coupling $ W\widetilde\chi _1^0
\widetilde\chi _1^{\pm}$ is much larger than that of $
Z\widetilde\chi _1^0 \widetilde\chi _1^0$. Note that annihilation
rate of the latter one is proportional to $ \left| {N_{14} }
\right|^2-\left| {N_{13} } \right|^2$.  In Fig. \ref{hig}, we have
shown that $ \left| {N_{14} } \right|^2 + \left| {N_{13} }
\right|^2$ is small over most of the parameter space, thus the
coupling $Z\widetilde\chi _1^0 \widetilde\chi _1^0$ is usually
small. Especially in the vicinity of EWSB boundary, in which $
\widetilde\chi _0^1$ can have large component of Higgsino, the
fraction of $ \widetilde\chi _0^1$ annihilation to $ W^ + W^ -$ is
always much more larger than $ 90\%$. In Fig. \ref{br}, the
fractions of channel $ W^ +  W^ - , ZZ, Zh, t\overline t $ are
plotted. From the figure we can see that the dominant annihilation
channel is $ W^ + W^ -$. For other channels $ZZ, Zh, t\overline t $,
only very few parameter points can have fraction above 10$\%$.

The energy distribution of induced neutrino and anti-neutrino from $
\widetilde\chi _0^1$ annihilation is the same, thus we can only
consider the neutrinos. Because we plan to investigate the neutrino
observation by IceCube which has the threshold energy around 50 GeV,
we will focus on the flux of high energy neutrino with energy
greater than 50 GeV. The production of high energy $ \nu_\mu ,
\nu_e$ neutrinos are the same. The reason is that the produced $\mu$
is stopped by the matter in the Sun before its decay \cite{RS88},
thus it mainly produces neutrinos with low energy which we do not
consider in this paper. Neutrinos can also be produced from charged
pions from hadronization of partons, but they contribute mainly to
low energy spectrum. Therefore for high energy neutrinos they are
small compared with those from gauge boson and top quark decays. The
production of $ \nu _\tau$ are usually larger than $ \nu _\mu ,\nu
_e$ because $ \tau$ lepton decays before it loses energy
\cite{CFMSSV05}. In summary the neutrino flavor ratios are given by
\cite{CFMSSV05}
\begin{equation}
v_e :v_u :v_\tau  :\overline v _e :\overline v _u :\overline v _\tau
= 1:1:r:1:1:r,
\end{equation}
where $r$ varies with the energy of neutrino.

We utilize Calchep \cite{comphep,calchep} to generate events, and
then use Pythia \cite{pythia} to handle the decays of intermediate
particles contained in the events.

\subsubsection{Annihilation channel $\widetilde\chi _0^1
\widetilde\chi _0^1  \to W^ +  W^ -,ZZ$ }

For the annihilation channel $ \widetilde\chi _0^1 \widetilde\chi
_0^1 \to W^ +  W^ -,ZZ$, the polarized effects of gauge bosons
should be considered, which makes the energy distribution of final
state neutrinos different from those without including polarization
information \cite{BKST07}. In this channel, gauge bosons are
transverse polarized with equal up- and down-type polarizations.

\begin{figure}[h]
\vspace*{-.03in} \centering
\includegraphics[width=3.3in,angle=0]{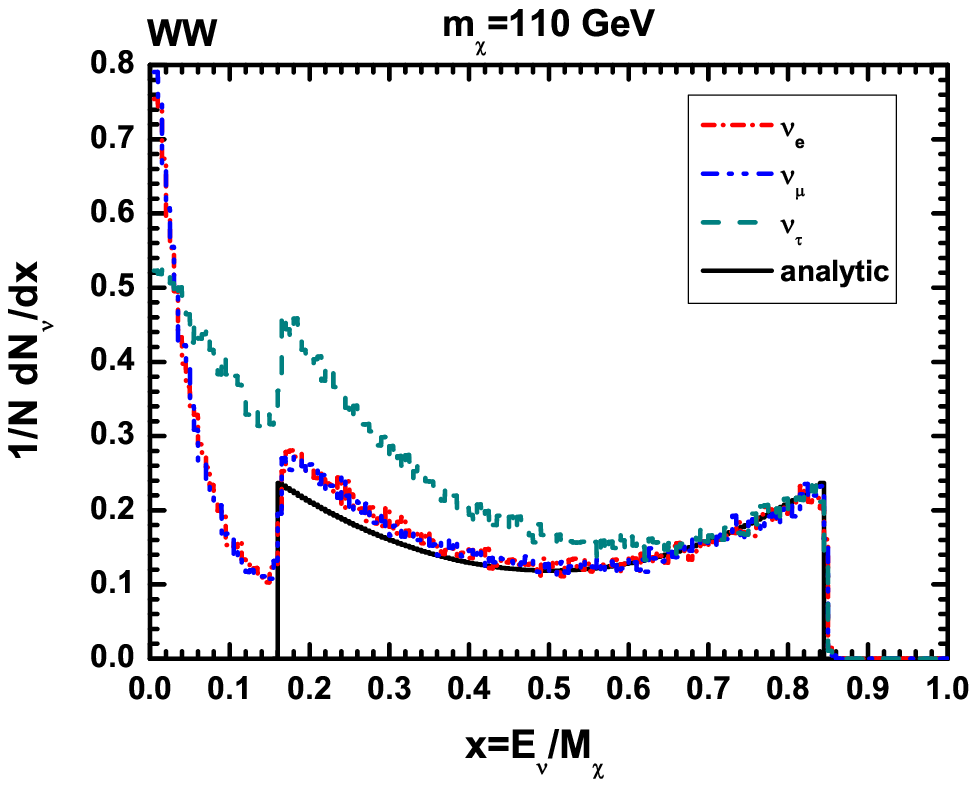}%
\includegraphics[width=3.3in,angle=0]{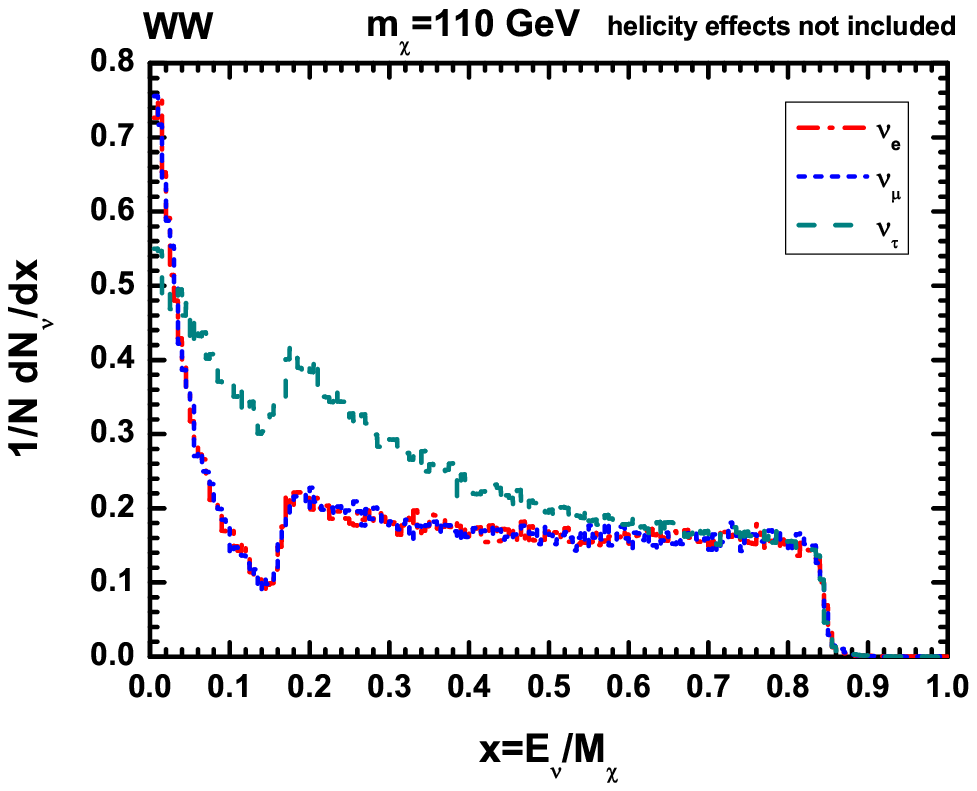}
\\
\includegraphics[width=3.3in,angle=0]{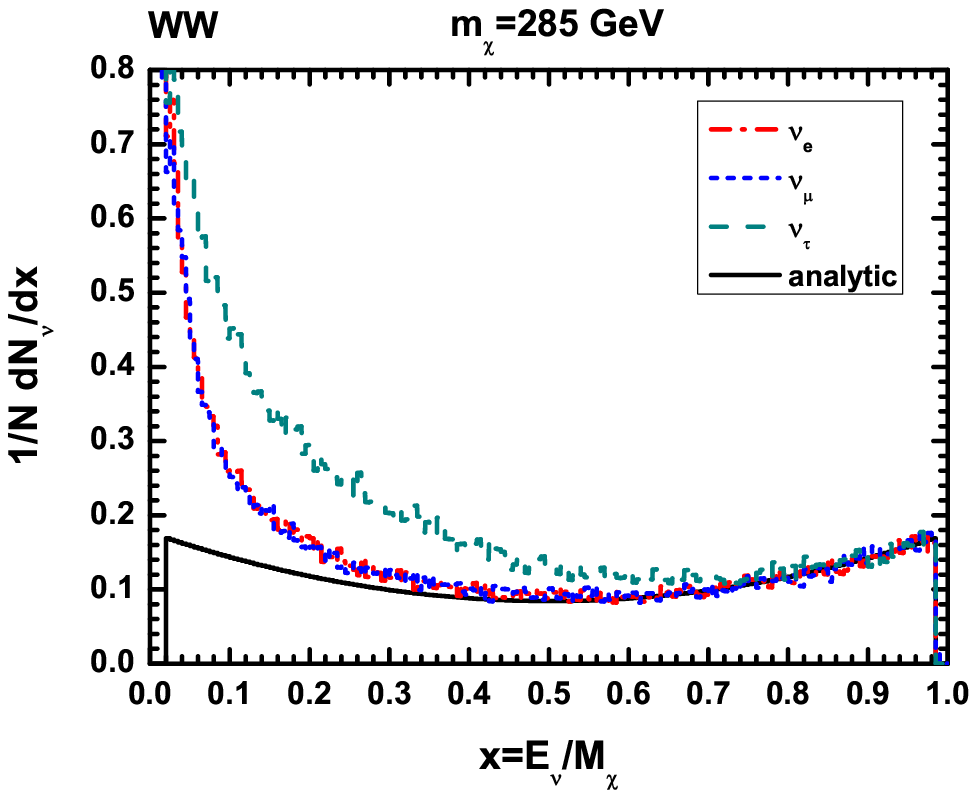}%
\includegraphics[width=3.3in,angle=0]{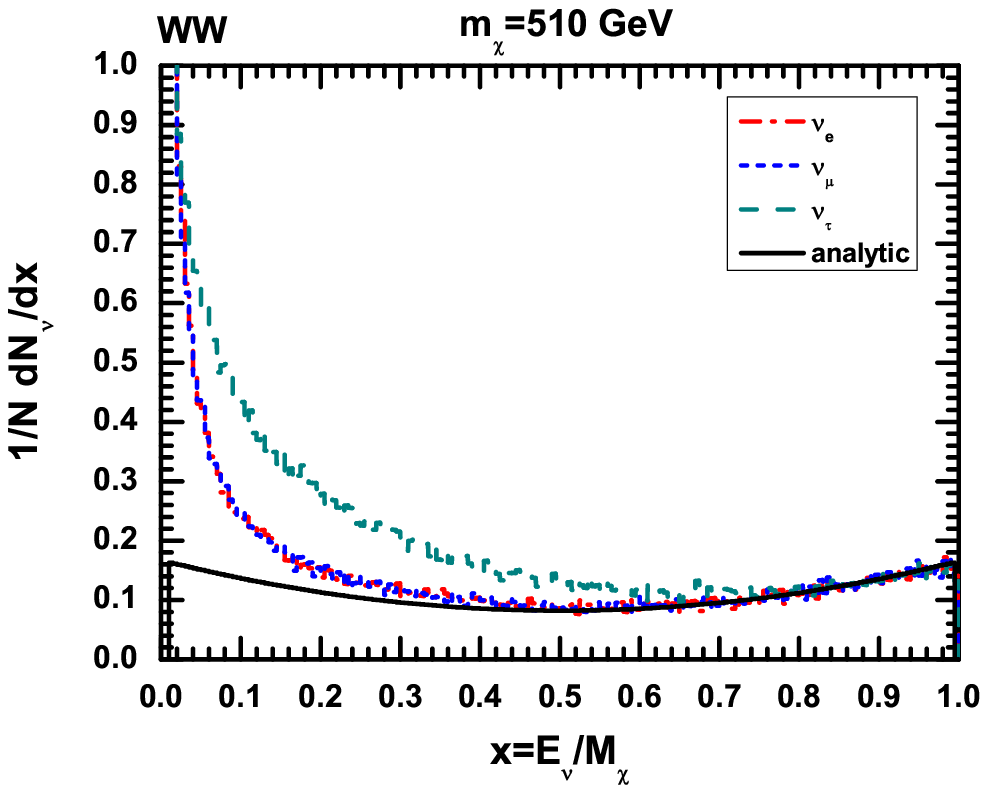}
\vspace*{-.03in} \caption{Energy distributions of neutrinos from
annihilation channel $ \widetilde\chi _0^1 \widetilde\chi _0^1 \to
W^ +  W^ -$ with neutralino mass 110, 285 and 510 GeV respectively.
For the right-top plot, the polarization effect of W gauge bosons
are not included. The horizontal axis is energy $ x = \frac{{E_\nu
}}{{m_{\widetilde\chi _0^1 } }}$. The vertical axis is $
\frac{1}{N}\frac{{dN_\nu }}{{dx}}$, where N is the number of
annihilation $ \widetilde\chi _0^1 \widetilde\chi _0^1 \to W^ + W^
-$ and $ {N_\nu  }$ is the number of neutrinos. The dashed-dot
lines, the short dashed lines and the dash lines represent $ \nu _e
,\nu _\mu ,\nu _\tau$ neutrinos respectively. The solid lines stand
for the analytical results of neutrinos which come directly from
leptonic decay of polarized W gauge boson  produced in $
\widetilde\chi _0^1 \widetilde\chi _0^1 \to W^ +  W^ -$
\cite{BKST07}. \vspace*{-.1in}} \label{wwprod}
\end{figure}

In Fig. \ref{wwprod}, the energy distributions of neutrino are shown
for several typical neutralino masses. It is obvious that the
polarization effects can change the shape of the energy
distributions. In the figure the solid lines in the bottom are the
analytical results of neutrinos which come directly from W gauge
boson leptonic decay \cite{BKST07}. The dashed-dot lines in the
middle are electron neutrino. We can see that the monte carlo
results are higher than analytical results because we include the
contributions from varies particle decays. Another distinction
between analytical results and monte carlo results is that the
latter ones have a higher tail in the low energy region because of W
gauge boson hadronic decay and tau lepton decay.
 The dash lines on the top are the tau neutrinos, which are higher
 than those of electron neutrinos duo to the contributions from
 tau lepton decay. The
muon neutrino has the same distribution as electron neutrino. Note
that in the leptonic decay of W gauge boson, the muon is set to be
stable in our simulations, i.e. the neutrino produced by the muon is
not counted. The reason is that the $\mu$ loses energy quickly in
the Sun and almost at rest before it decays into neutrino
\cite{RS88}. The decay of tau lepton has significant contributions
to tau neutrino compared with analytical result which only include
neutrino from W boson \cite{CFMSSV05,BEO07}. It will enhance the
number of muon neutrino via the oscillation effect \cite{BEO07}.

\begin{figure}[h]
\vspace*{-.03in} \centering
\includegraphics[width=3.3in,angle=0]{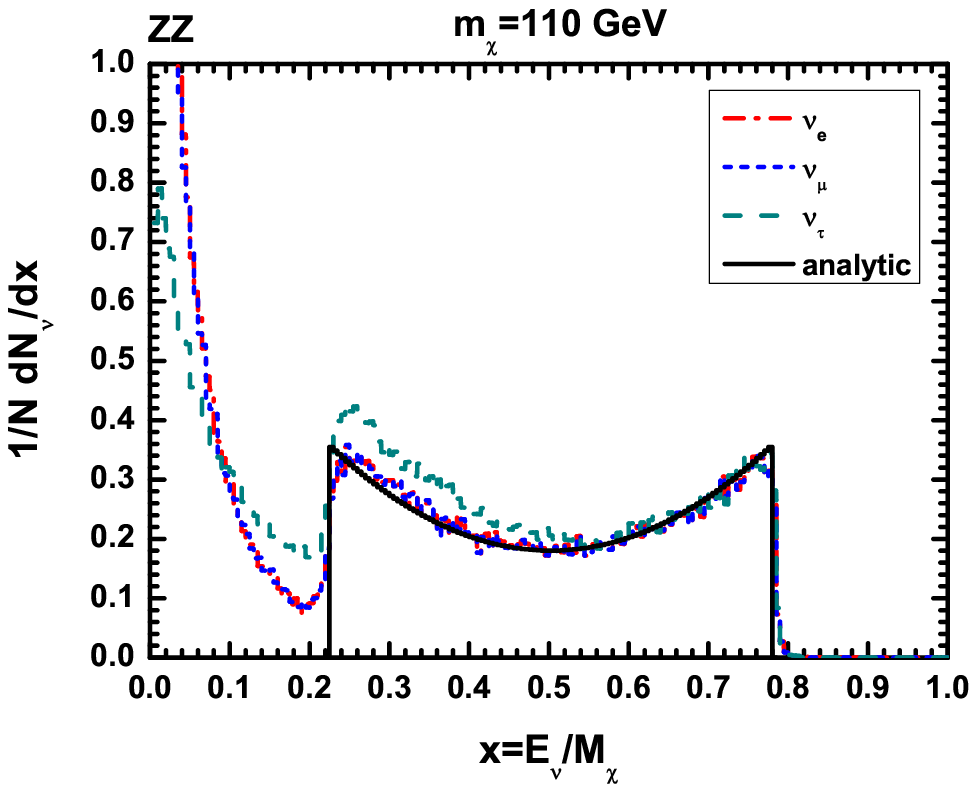}%
\includegraphics[width=3.3in,angle=0]{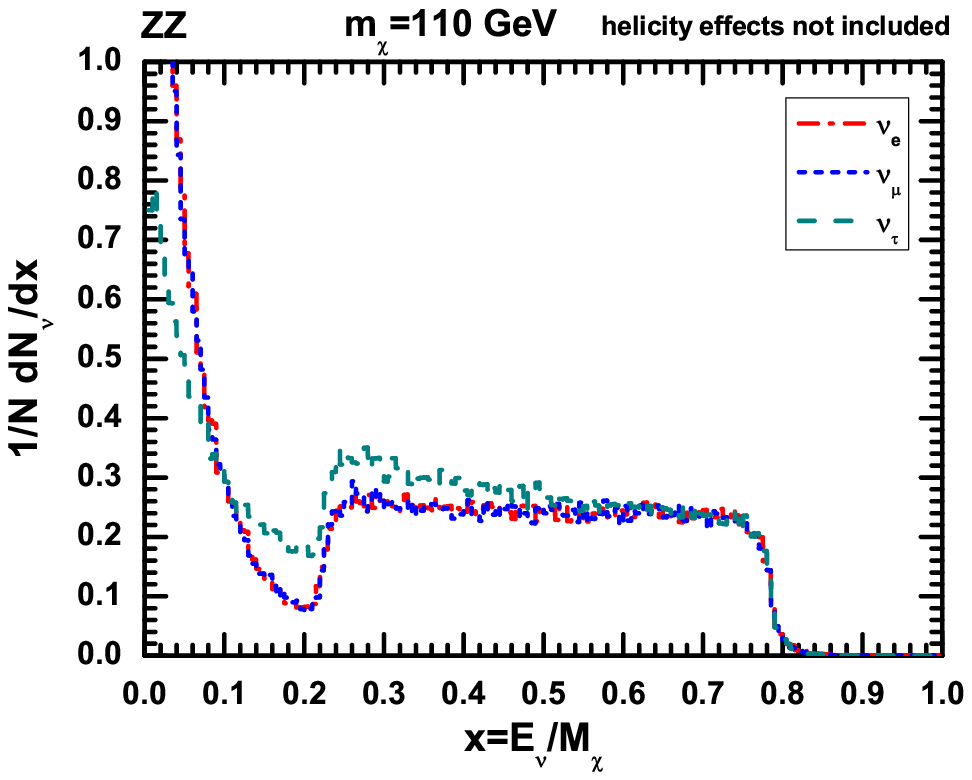}
\vspace*{-.03in} \caption{ Same as Fig.~\ref{wwprod} but for $
\widetilde\chi _0^1 \widetilde\chi _0^1 \to Z Z$.
 \vspace*{-.1in}}
\label{zzprod}
\end{figure}

In Fig.~\ref{zzprod}, the energy distributions of neutrinos from
annihilation channel $ \widetilde\chi _0^1 \widetilde\chi _0^1 \to Z
Z$ are plotted. For 110 GeV neutralino, the ZZ channel has narrower
energy distribution compare to that of WW because Z is heavier than
W.   For heavier neutralino such different energy distribution
between Z and W will be smaller. $ \frac{1}{N}\frac{{dN_\nu
}}{{dx}}$ in ZZ channel is higher than that
 in WW channel because there are two Z bosons and each Z can contribute to
both neutrino and anti-neutrino while W boson can not.

\subsubsection{Annihilation channel $\widetilde\chi _0^1 \widetilde\chi _0^1  \to Z h$ }

\begin{figure}[h]
\vspace*{-.03in} \centering
\includegraphics[width=3.3in,angle=0]{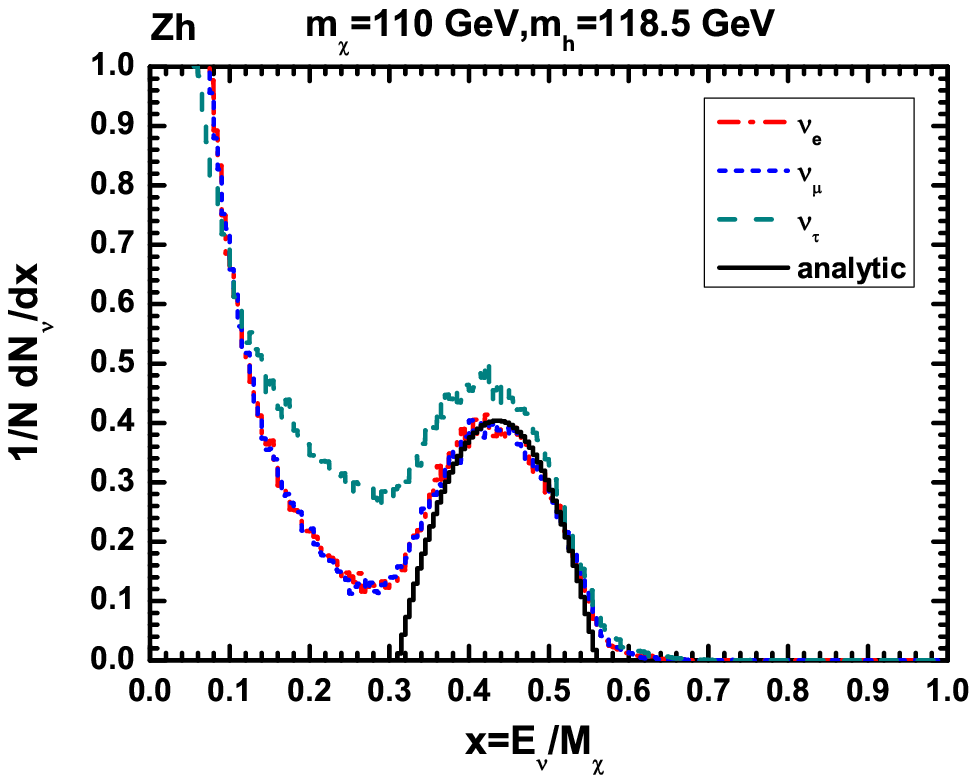}%
\includegraphics[width=3.3in,angle=0]{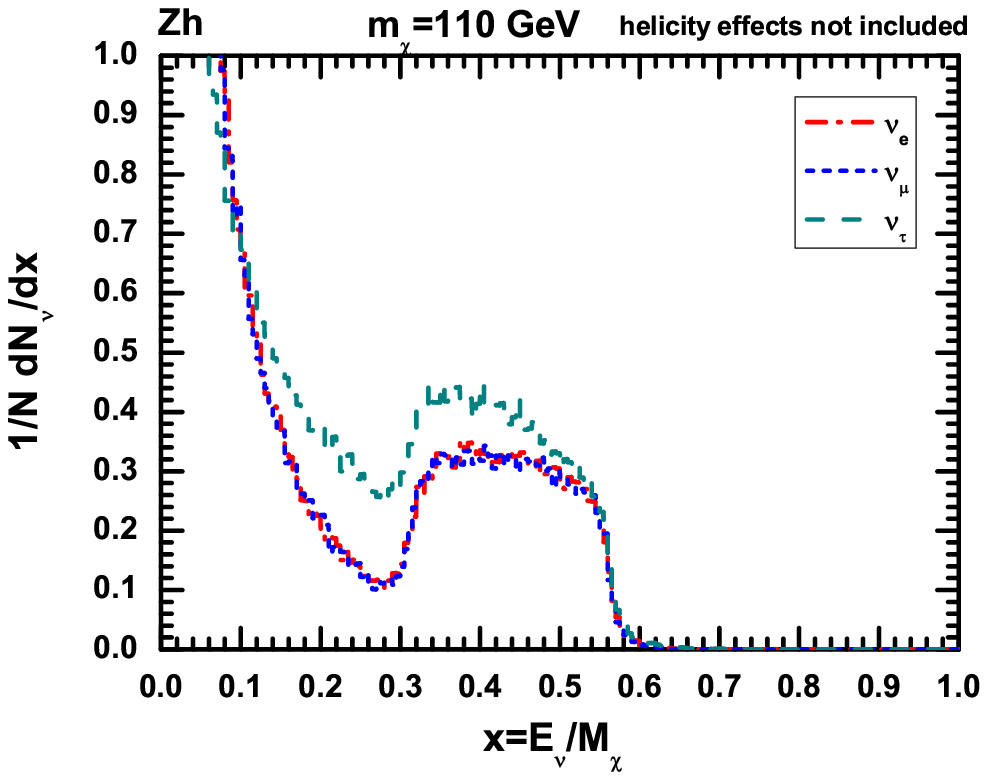}
\\
\includegraphics[width=3.3in,angle=0]{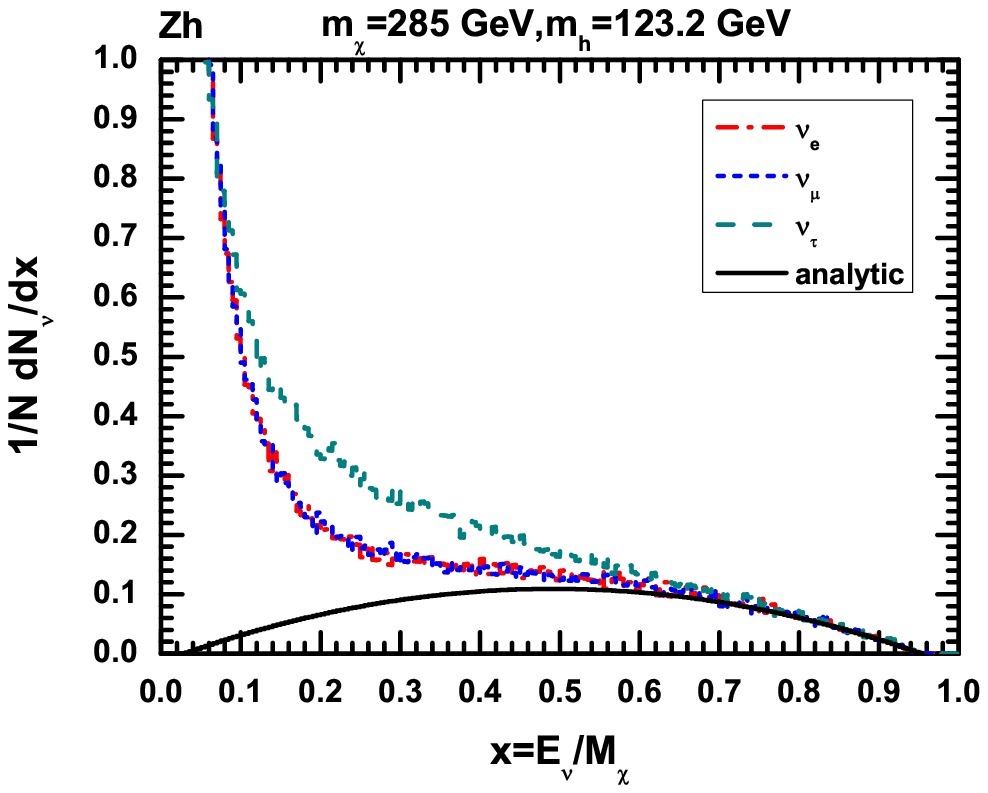}%
\includegraphics[width=3.3in,angle=0]{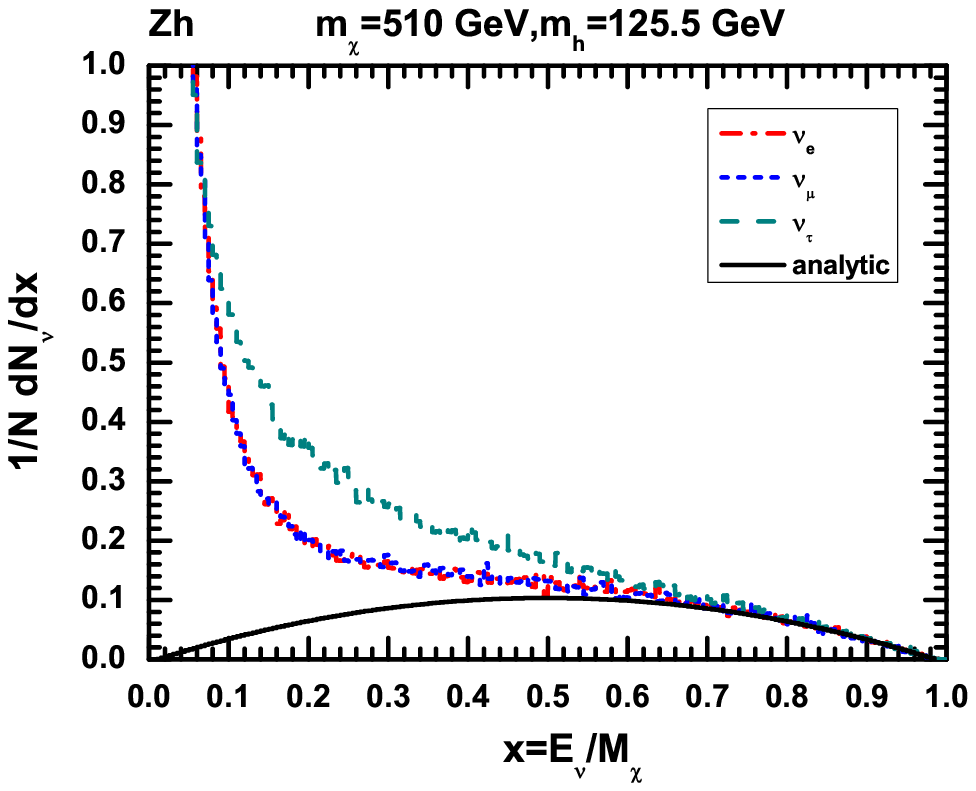}
\vspace*{-.03in} \caption{Same as Fig.~\ref{wwprod} but for
$\widetilde\chi _0^1 \widetilde\chi _0^1 \to Z h$.
 \vspace*{-.1in}}
\label{zhprod}
\end{figure}

In annihilation channel $\widetilde\chi _0^1 \widetilde\chi _0^1 \to
Z h$, the Z boson is longitudinal polarized \cite{BKST07}. The
energy distributions of neutrinos are presented in
Fig.~\ref{zhprod}. We also compared the analytic results with our
full simulations which include both polarization effects and various
particle decays. Once again, the full simulations have higher tau
neutrino production, right energy distributions shape after
including polarization effects, as well as the higher low energy
tail from lepton decay and quark hadronic decays.

\subsubsection{Annihilation channel $\widetilde\chi _0^1 \widetilde\chi _0^1  \to t\overline
t$ }

\begin{figure}[h]
\vspace*{-.03in} \centering
\includegraphics[width=3.3in,angle=0]{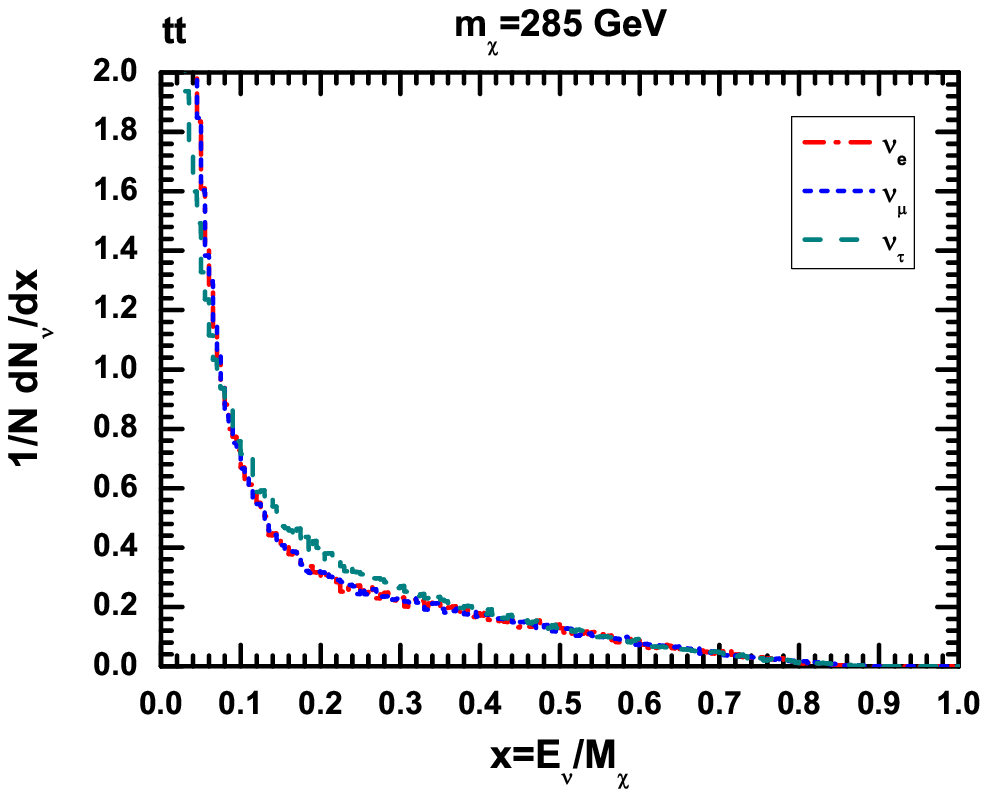}%
\includegraphics[width=3.3in,angle=0]{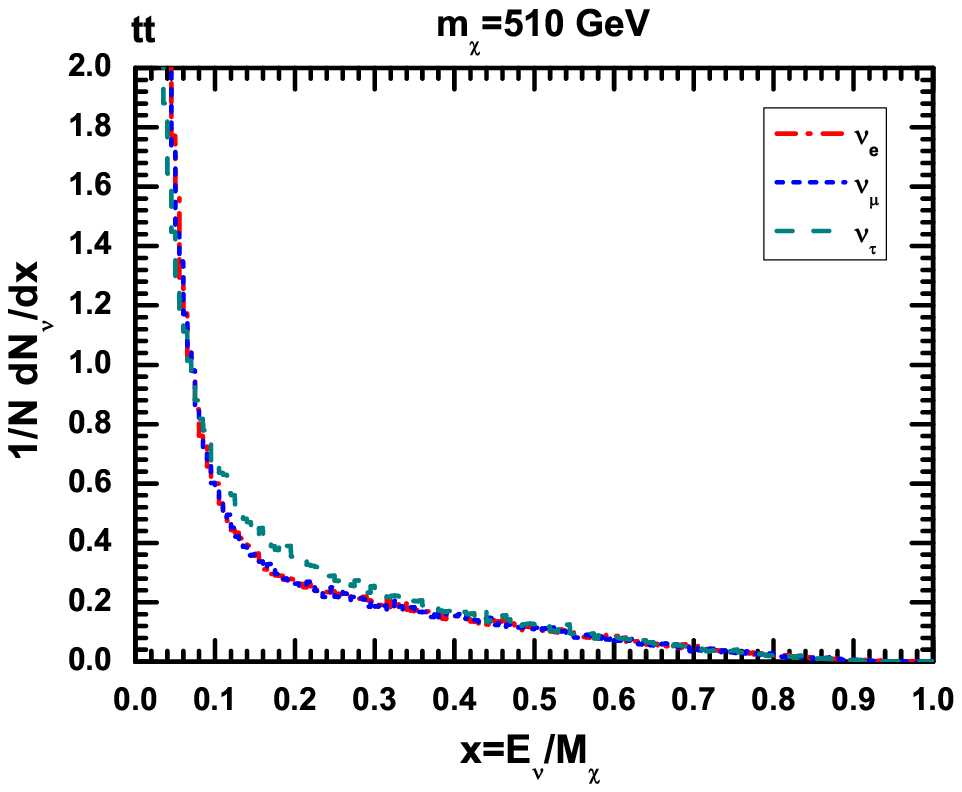}
\vspace*{-.03in} \caption{Energy distributions of neutrinos from
annihilation channel $\widetilde\chi _0^1 \widetilde\chi _0^1  \to
t\overline t$ with neutralino mass 285 and 510GeV respectively. The
conventions are the same as those in Fig.~\ref{wwprod} .
\vspace*{-.1in}} \label{ttprod}
\end{figure}

For annihilation channel $\widetilde\chi _0^1 \widetilde\chi _0^1
\to t\overline t$, it has small contributions to our final results.
For simplicity we utilize WimpSim to generate the signals with
averaged t-quark polarizations \cite{wimpsim}. The energy
distributions are shown in Fig.~\ref{ttprod}.

\subsection{The propagation of neutrino}

The neutrinos produced at the solar center will interact with the
nucleus in the Sun during their propagation to the solar surface.
The matter effects include the neutral current(NC) effects, the
charged current(CC) effects and tau neutrino $ \nu _\tau$
re-injection from secondary tau lepton decay. From solar center to
the Earth the oscillation effects of neutrinos are also important
and must be taken into account. In the density matrix method, the
evolution equation is \cite{CFMSSV05}
\begin{equation}
\frac{{d\rho }}{{dr}} =  - i[H,\rho ] + \left. {\frac{{d\rho
}}{{dr}}} \right|_{NC}  + \left. {\frac{{d\rho }}{{dr}}}
\right|_{CC}  + \left. {\frac{{d\rho }}{{dr}}} \right|_{0}.
\label{prop}
\end{equation}
Here the first term at the RHS describes the vacuum oscillation. The
second and third terms describe NC and CC effect separately,
including neutrino absorption, scattering and $ \nu _\tau$
re-injection. The last term $\left. {\frac{{d\rho }}{{dr}}}
\right|_{0}=\delta(r)\delta_{ij}\frac{1}{N}\frac{dN}{dE_{\nu}}$ is
the initial neutrino at the production point.

We use Monte Carlo program WimpSim which is event-based code to
handle the propagation process of neutrinos from the Sun center to a
distance of 1 astronomical unit (AU) \cite{wimpsim}. WimpSim treats
three flavor neutrino interactions and oscillations in one
framework.  The equivalence between the Monte Carlo method and the
density matrix formalism used in \cite{CFMSSV05} has been discussed
in Ref. \cite{BEO07}. The detail Monte Carlo simulations allow us
tracing every neutrino event and determining final $\mu$ signals
with the concrete energy and angle information. Thus this approach
can provide more realistic events. We take the neutrino oscillation
parameters which given by global fit to neutrino experiments
\cite{MSTV07} ( The 2007 updated values in Ref. \cite{MSTV07} have
been used ). Explicitly, we choose $\theta _{{\rm{12}}} = 34.4^
\circ$, $ \theta _{{\rm{13}}} = 0^ \circ$ and $ \theta _{{\rm{23}}}
= 45^ \circ$, the mass-squared difference   $ \Delta m_{21}^2 = 7.6
\times 10^{ - 5} eV^2$, $ \Delta m_{31}^2  = 2.4 \times 10^{ - 3}
eV^2$ and CP-violating phase $\delta=0$.

\section{The detection of neutrino by IceCube}

When the Sun is below the horizon, the up-going muon-neutrinos may
produce high energy muons in the Earth. When these muons travel in
the ice, they will lose their energy via scattering with other
matter. The IceCube in south pole is designed to detect such kind of
muons.

The analytical formula of final muon flux at the detector has been
given in Ref. \cite{BKST07}. The muon energy loss is given as
\cite{DRSS00}
\begin{equation}
\frac{dE}{dx}=-\alpha-\beta E,
\end{equation}
where $\alpha$ and $\beta$ are empirical parameters.
 The distance that a muon travels in the Earth before its
energy drops below threshold energy $E^{thr}$, is called muon range
which is given as
\begin{equation}
R_{\mu}(E)=\frac{1}{\rho\beta}\ln(\frac{\alpha+\beta E}{\alpha+\beta
E^{thr}}).
\end{equation}

Instead of utilizing above analytical formulas to calculate muon
flux, we use WimpSim to get muon flux based on  event-by-event Monte
Carlo simulation. After obtaining the neutrino events at the surface
of the Earth as discussed in the previous section, we then simulate
the neutrinos propagation in the Earth and the final flux of muons.
The muons are induced from CC interaction at actual neutrino
telescope IceCube, which is mainly characterized by $E^{thr}$,
$R_{\mu}(E)$, $A_{eff}$ and its geographical position. We calculate
the effective detecting area $A_{eff}$ as that in Ref. \cite{GHM05}.
Note that $A_{eff}$ in Eqn. \ref{eq0} is also function of final muon
energy. The WimpEvent uses a simple model to treat the Earth orbit
and obtain the detector time-dependent location for each neutrino
event \cite{BEO07}.

Comparing with analytical formula, the simulation can give more
information about muon flux. The muon energy at the detector can be
given by this simulation, while analytical formula only gives muon
energy at the vertex where neutrinos convert into muons. The
information about angle between muon velocity and the direction from
the Sun to the detector can also be obtained which is important to
reduce background.

\begin{figure}[h]
\vspace*{-.03in} \centering
\includegraphics[width=4.0in,angle=0]{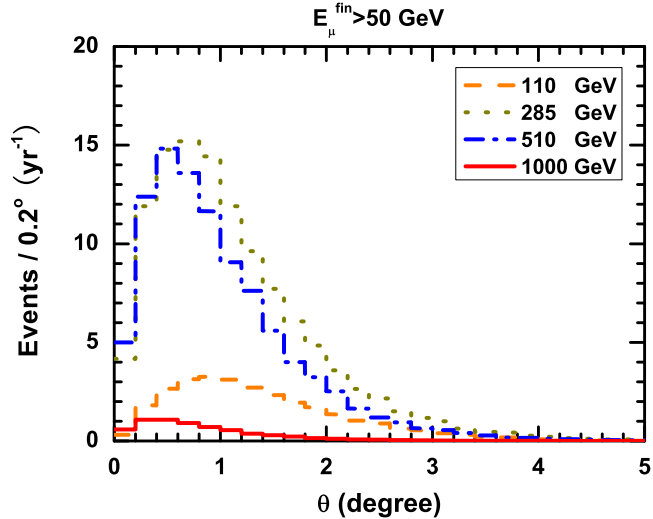}
\vspace*{-.03in} \caption{The muon events ($\mu^++\mu^-$)
distributions in IceCube as a function of  angle $\theta$ which is
defined as angle between the muon velocity and the direction from
Sun to the detector. The IceCube $1^0$ angle resolution is not
reflected here, and every bin is taken as $0.2^0$. The threshold
energy for muon is taken as $50 GeV$. \vspace*{-.1in}}\label{angle}
\end{figure}

We show our final results for some benchmark points in Fig.
\ref{angle} after multiplying initial dark matter annihilation rate
in the core of the Sun. Here we have summed the $\mu^{\pm}$ events
together and taken $E^{thr}=50GeV$ into account. After produced from
charged current scattering process, the muon may have multiple
Coulomb scattering in the ice \cite{E93,EG95}. Thus the final
direction of muon velocity is not the exactly same direction from
the Sun to the detector. From Fig. \ref{angle}, we can see the
deviation of muon direction is not large, and most of muon events
are in the scope of $\pm 2^0$ around the direction from the Sun to
the detector. And the signals from high energy neutrinos produced by
heavy massive neutralinos annihilation are less affected by Coulomb
scattering as naive expectation. Note that the IceCube  can reach
angle resolution as small as $1^0$ \cite{Aetal04}, our results
suggest a possible method to reduce the background events from other
sources.

Now we switch to the discussions on the backgrounds. The largest
backgrounds come from comic ray. When cosmic ray interacts in the
atmosphere around the Earth, high energy neutrinos would be created
and travel to the detector. During the propagation, the neutrinos
can produce muons \cite{atmnu,E93,HKKMS07}. Such kind of high energy
neutrinos do not have special direction. Therefore if we only
observe the events around the special direction from Sun to the
detector, the backgrounds would be dramatically reduced. Other
possible background are muons from cosmic ray
 interacting with the atmosphere around the Earth  \cite{atmmu} and
neutrinos from cosmic ray interacting with particles in the Sun's
corona \cite{suncorona}.  However these two kinds of backgrounds are
not important in the detection of muons at IceCube. For the rough
estimate, we only take the atmosphere neutrino backgrounds into
account.  The atmosphere neutrino flux are taken from Ref.
\cite{HKKMS07} and method to estimate the muon rate from Ref.
\cite{BKST07}. The resulting background muons are not the function
of $E^{fin}$ but for $E^0$, i.e. energy of the induced muon.  It
should be noted that more accurate results need Monte Carlo
simulation.

\begin{figure}[h]
\vspace*{-.03in} \centering
\includegraphics[width=3.3in,angle=0]{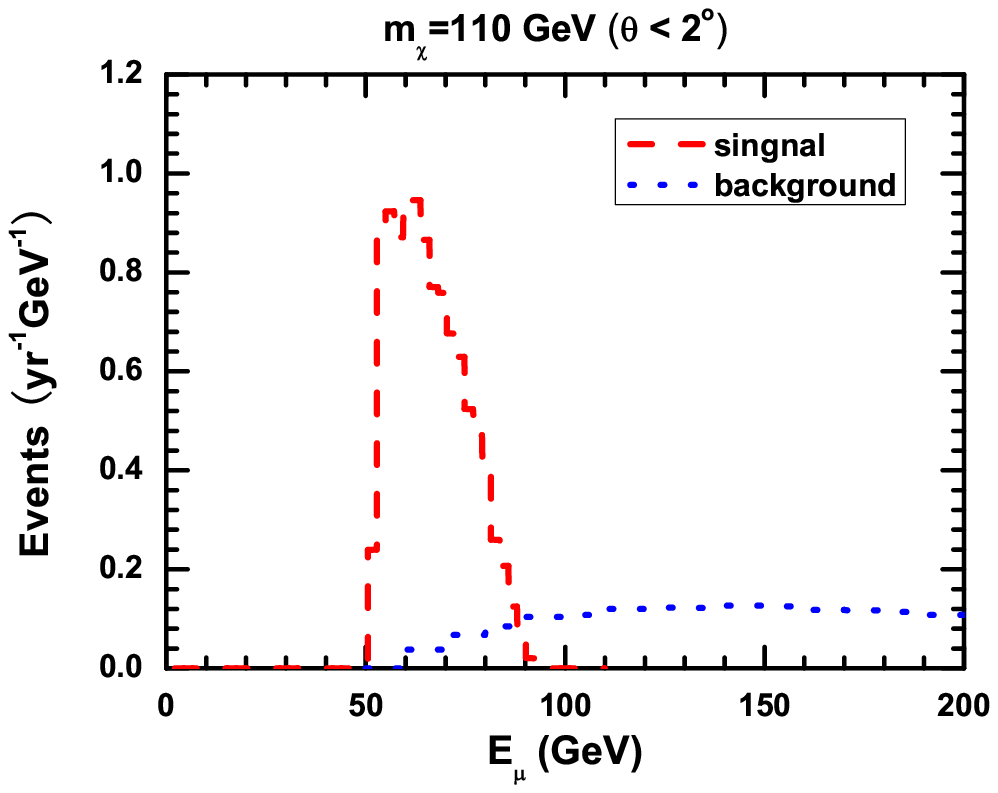}%
\includegraphics[width=3.3in,angle=0]{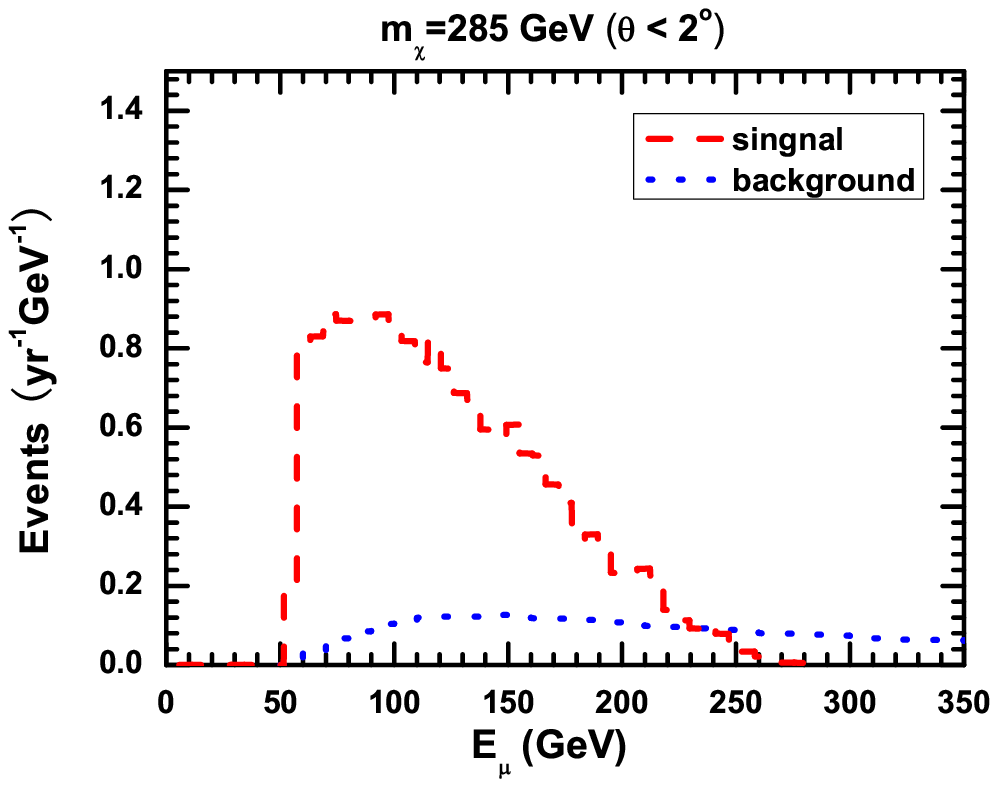}
\\
\includegraphics[width=3.3in,angle=0]{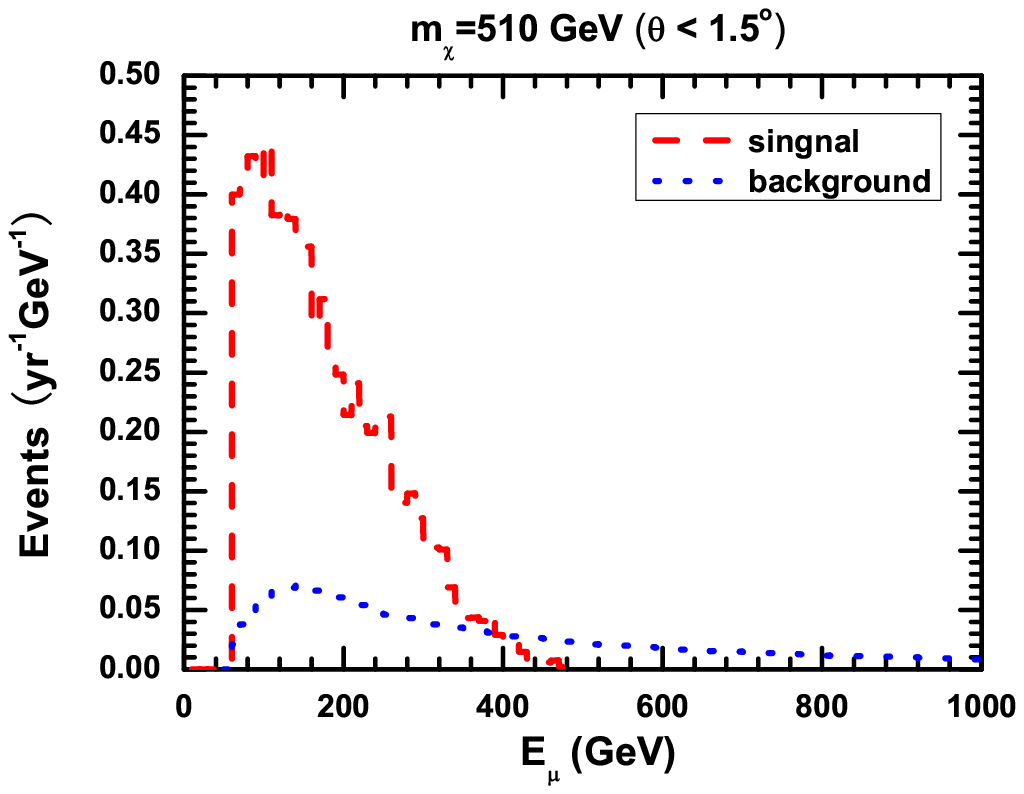}%
\includegraphics[width=3.3in,angle=0]{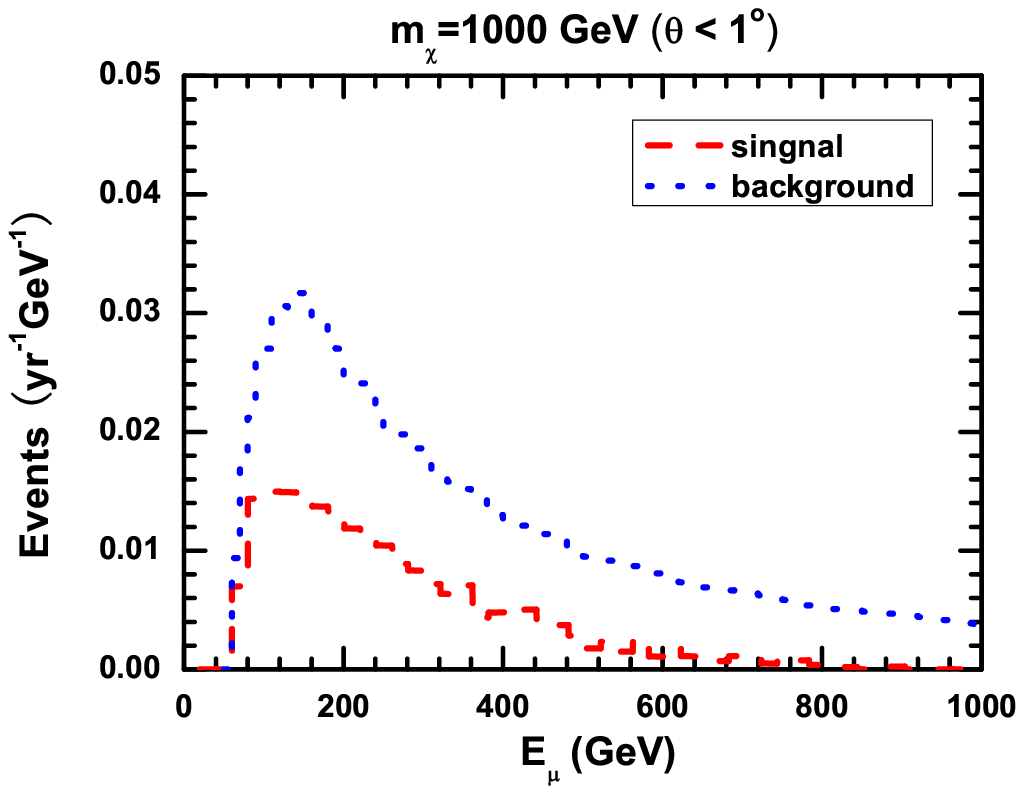}
\vspace*{-.03in} \caption{Differential muon rate in IceCube as a
function of energy of muon for several neutralino mass as 110, 285,
510 and 1000 GeV. We applied different angular cuts on muon zenith
angle in order to get better significance  according to the result
of Fig.~\ref{angle}. The red dashed (the blue dotted ) lines
represent the muon signals (the atmosphere backgrounds).
 \vspace*{-.1in}}
\label{finmu}
\end{figure}

\begin{table}[htb]
\begin{tabular}{||c||c|c|c|c||c|c|c||}
\hline \hline
 Point & $m_\chi$(GeV) & $m_0$(TeV) & $m_{3/2}$(TeV) & $\mu$(GeV) & $R_{h}$ &  $\sigma_{SD}(pb)$ &
 $ C_ \odot  (yr^{ - 1} ) $   \\
\hline
A & $ 110$ &\,\,2.9 &\,\,34 &\,\,420 &\,\, 4.4 $\%$&\,\, 5.1$\times 10^{ - 5}$&\,\,8.3$\times 10^{ 29}$ \\
\hline
B & $285$ &\,\,6.1 &\,\,88 &\,\,498&\,\,7.1 $\%$&\,\,4.2$\times 10^{ - 5}$&\,\,1.0$\times 10^{ 29}$\\
\hline
C& $510$ &\,\,9.5 &\,\,162 &\,\,620 &\,\,19.2 $\%$&\,\,5.0$\times 10^{ - 5}$&\,\,4.1$\times 10^{ 28}$\\
\hline
D & $1000$ &\,\,15.3 &\,\,314 &\,\,1120 &\,\,14.5 $\%$&\,\,1.1$\times 10^{ - 5}$&\,\,2.2$\times 10^{ 27}$\\
\hline \hline
\end{tabular}
\caption{Several benchmark points (point A-D) in mAMSB model. $ R_h$
is the higgsino fraction $ \left| {N_{13} } \right|^2 + \left|
{N_{14} } \right|^2$, $ \sigma _{SD}$ is the spin-dependent cross
section and $ C_ \odot$ is the neutralino capture rate for the Sun.
 \vspace*{-.1in}}\label{table1}
\end{table}

\begin{table}[htb]
\begin{tabular}{||c||c|c|c|c||c|c|c|c||}
\hline \hline Point & $\frac{{\sigma _{WW} }}{{\sigma _{tot}
 }}$&
 $\frac{{\sigma _{ZZ} }}{{\sigma _{tot} }}$&
 $\frac{{\sigma _{t\overline t } }}{{\sigma _{tot} }}$&
 $
\frac{{\sigma _{Zh} }}{{\sigma _{tot} }}$
 &$\theta cut$  ($^\circ$)& $Sig$ & $ BG $& $\sigma _{stat}$   \\
\hline
A& \,\,100.0$\%$&\,\,0.0$\%$&\,\,0.0$\%$&\,\,0.0$\%$&\,\,2.0&\,\,23.3&\,\,24.1&\,\,4.7\\
\hline
B &\,\,99.1$\%$&\,\,0.0$\%$&\,\,0.9$\%$&\,\,0.0$\%$&\,\,2.0&\,\,102.1&\,\,24.1&\,\,20.8\\
\hline
C &\,\,95.6$\%$&\,\,0.2$\%$&\,\,4.2$\%$&\,\,0.0$\%$&\,\,1.5&\,\,70.3&\,\,13.6&\,\,19.1\\
\hline
D &\,\,96.9$\%$&\,\,0.1$\%$&\,\,3.0$\%$&\,\,0.0$\%$&\,\,1.0&\,\,3.0&\,\,6.0&\,\,1.2\\
\hline \hline
\end{tabular}
\caption{One year total muon rate at IceCube for various benchmark
points (point A-D) where muons  satisfy $ E_{thr}  < E_\mu < 300GeV$
with $ E_{thr} = 50GeV$. Here $ \frac{{\sigma _{WW} }}{{\sigma
_{tot} }},\frac{{\sigma _{ZZ} }}{{\sigma _{tot} }},\frac{{\sigma
_{t\overline t } }}{{\sigma _{tot} }},\frac{{\sigma _{Zh} }}{{\sigma
_{tot} }}$ are the fractions of annihilation channels $
WW,ZZ,Zh,t\overline t$ separately. $\theta _{CUT}$ is the angular
cut for zenith angle of muon. Sig is the annual muon rate and BG is
the annual muon background from atmosphere neutrino. We define
statistical significance as $ \sigma _{stat} = S/\sqrt B $.
 \vspace*{-.1in}}\label{table2}
\end{table}

Our final results are depicted in Fig.~\ref{finmu},
Tab.~\ref{table1} and \ref{table2}. The four parameter points are
almost dominant by annihilation channel WW which is one of the
characters of AMSB model. All four benchmark points, which satisfy
all the constraints from other experiments, have the same order of
magnitude of $ \sigma _{SD}$, which is essential for the muon
detection at IceCube. Point A with 110GeV neutralino has much
smaller signal than that of Point B with $285GeV$ neutralino because
the former is affected by the energy threshold and muon range
$R_\mu$. Point C and D with heavier neutralino are less affected by
the energy threshold but the capture rate $C_ \odot$ is suppressed.
We applied different angular cut on zenith angle of muon to get
better statistical significance according to the results in
Fig.~\ref{angle}. From Tab.~\ref{table2}, we can see the signal
annual events can reach 102 for Point B and  the statistical
significance $ \sigma _{stat}$ can reach more than 20. Based on our
numerical estimation, we can obtain the rule of thumb for IceCube to
discover high energy neutrino from Sun in AMSB model, i.e.  $ \sigma
_{SD}
> 10^{ - 5} pb$. It should be noted that $ N_{13}^2  + N_{14}^2  > 4\%$ approximately
corresponds to the condition $ \sigma _{SD}  > 10^{ - 5} pb$ based
on the results in Fig.~\ref{relic} and \ref{sdvall}.

\section{Conclusions and discussions}

In this paper in the mAMSB model we have investigated the muon event
rate at IceCube. Such kind of muon events are induced by high energy
neutrino flux from neutralino annihilation in the Sun. We studied
the detail energy and angular spectrum of the final muons at the
detector based on event-by-event simulation by using Monte Carlo
code WimpSim. More precisely we simulated the processes since the
production of neutrino via neutralino annihilation in the core of
the Sun, neutrino propagation from the Sun to the Earth, as well as
the converting processes from neutrino to muon. Our results show
that in the mAMSB model it is possible to observe the energetic
muons at IceCube, provided that the lightest neutralio has
relatively large higgsino component, as a rule of thumb $ N_{13}^2 +
N_{14}^2  > 4\%$ or equivalently $\sigma _{SD}> 10^{ - 5} pb$.
Especially for our benchmark Point B (see Tab.~\ref{table1}), the
signal annual events can reach 102  and the statistical significance
can reach more than 20. This parameter space is very similar to the
'focus point' region in mSUGRA model.

It should be emphasized that our study has much more improvement
compared to previous investigations. For example, we include both
polarized effect of gauge bosons and secondary neutrinos from the
decay of leptons and quarks. The polarization effects of the gauge
bosons can have important influences on the neutrino energy
spectrum, as shown in our results and in Ref. \cite{BKST07}. The
secondary neutrinos are usually handled by Pythia, for example in
Ref. \cite{CFMSSV05} and WimpSim \cite{wimpsim}. But none of them
include both  effects. We use Monte Carlo code WimpSim to handle the
interactions includes NC effect, CC effect and tau neutrino $ \nu
_\tau$ re-injection from secondary tau lepton decay. Comparing with
the previous analytical result, the Monte Carlo approach can not
only give the same evolution results as those by solving
Eqn.~\ref{prop} but also the angular distribution of muons as shown
in Fig.~\ref{angle}. With the angular distributions at hand, we can
choose $ 2^ \circ$ as normal cut in range $ 100GeV <
m_{\widetilde\chi _0^1 } < 500GeV$ to reduce the background from
atmosphere neutrinos. With heavier neutralinos, smaller cut can be
chosen to further reduce the background. As shown in
Ref.~\cite{H08}, the light neutralino in this scenario is
disfavored. Thus the techniques to suppress background in order to
extract low signal events for heavy neutralino is very important.

Last but not least, we want to emphasize that the final energy
spectra of muons will have similar shape as those in
Fig.~\ref{finmu} in the mAMSB model, because the annihilation of
neutralinos are almost $ \widetilde\chi _0^1 \widetilde\chi _0^1 \to
W^ +  W^ -$. Thus it is possible to distinguish among AMSB model
from other SUSY breaking scenarios due to different annihilation
modes if enough muons can be collected.

\section{ Acknowledgements}

This work was supported in part by the Natural Sciences Foundation
of China (No. 10775001 and 10635030), and the trans-century fund of
Chinese Ministry of Education.

\end{document}